\documentstyle[11pt,twoside,colloqOHP,epsfig]{article}
\begin{document}
\title{The WHAT Project}
\author{A. Shporer$^1$, T. Mazeh$^1$, A. Moran$^1$, G. Bakos$^2$, G. Kovacs$^3$ \& E. Mashal$^1$}
\affil{$^1$Wise Observatory, Tel Aviv University\\ 
       $^2$Hubble Fellow, Harvard-Smithsonian Center for Astrophysics\\ 
       $^3$Konkoly Observatory, Hungarian Academy of Sciences} 
\begin{abstract}
We describe WHAT, a small-aperture short focal length automated 
telescope with an $8.2\deg \times 8.2\deg$ field of view, located at the Wise Observatory. The system is aimed at searching for transiting extrasolar planets and variable stars. Preliminary results of 3892 exposures of a single field are presented, where the telescope achieved already a precision of a few mmag for the brightest objects. Additional information can be found at: {\tt http://wise-obs.tau.ac.il/$\sim$what}.
\end{abstract}

\textbf{WHAT} is a \textbf{W}ise observatory \textbf{H}ungarian-made \textbf{A}utomated \textbf{T}elescope located at the Wise Observatory in Mizpe Ramon, Israel. It is a collaboration between the Wise Observatory of the Tel Aviv University and Konkoly Observatory of the Hungarian Academy of Sciences. Like all other HAT telescopes (Bakos et al. 2004, and references therein), WHAT is a combination of a fully automated telescope mount, a clamshell dome, 2K $\times$ 2K CCD, $200$ mm $f/1.8$ telephoto lens, RealTime Linux PC and a software environment, ``ProMount''.

WHAT operation is fully automated. The only manual interaction is that of establishing weather conditions and enabling the telescope on each evening. This nightly operation is done through a web interface, by a teamwork from Wise and Konkoly observatories.

The WHAT wide field of view, $8.2\deg \times 8.2\deg$, allows it to simultaneously observe tens of thousands of stars. Observations are carried out in a Point Spread Function (PSF) broadening mode, whereby the telescope is stepped on a prescribed pattern, thus achieving a better sampled PSF (Bakos et al. 2004). Most of WHAT observations are conducted through a standard Cousins $I$-band filter. For future calibration, a Johnson $V$-band filter is also used occasionally. Each exposure is $300$~s long, therefore, approximately $100$ object images are accumulated nightly.

WHAT primary objective is to search for transiting extrasolar planets by monitoring pre-selected fields. As a by-product, a large variable star inventory is also produced.
\begin{figure}[h]
\begin{center}
\epsfig{file=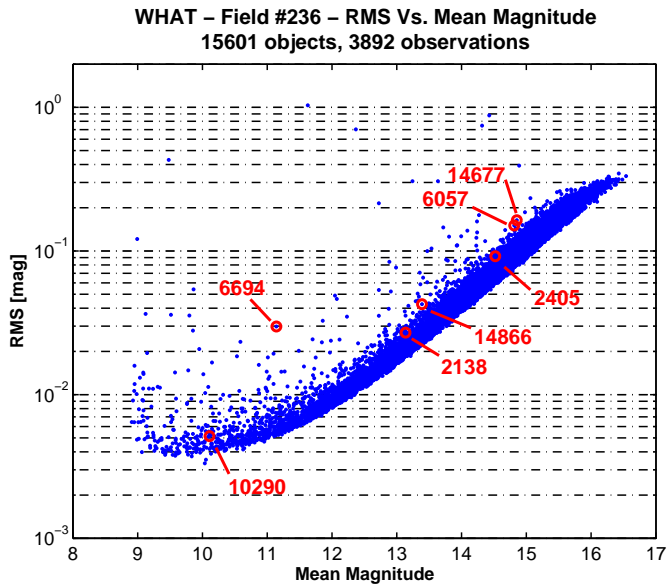,height=8cm,width=10cm}
\caption{Photometric results for WHAT field \#236. RMS is presented against instrumental $I$ magnitude, which is close to standard $V$. Objects presented in Figures \ref{fig:10290} and \ref{fig:lcs} are encircled here in red/grey, along with their ID numbers.}
\label{fig:RMS_mean}
\end{center}
\end{figure}

Over $65,000$ images have been accumulated since WHAT became operational on Jan. 2004. Preliminary results for one of the observed fields, \#236, are presented here. Located at R.A. 15:28 and Dec. +30:00, it was observed 3892 times on 162 nights between 2004 Feb 20 and 2004 Aug 1. Light curves of 15601 field objects were extracted using aperture photometry. RMS versus mean magnitude for all objects are presented in Figure \ref{fig:RMS_mean}. The systematic scatter reaches 4.5~mmag.

\begin{figure}[h]
\begin{center}
\epsfig{file=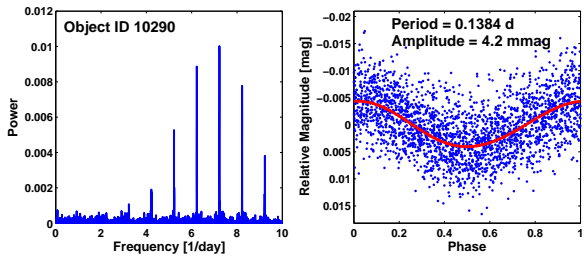,height=3.5cm,width=9cm}
\caption{\emph{Left:} Power spectrum of object 10290. 
\emph{Right:} Folded, mean subtracted, light curve, overplotted in red/grey by the corresponding Fourier fit, with an amplitude of 4.2~mmag.}
\label{fig:10290}
\end{center}
\end{figure}

\begin{figure}[h]
\begin{center}
\epsfig{file=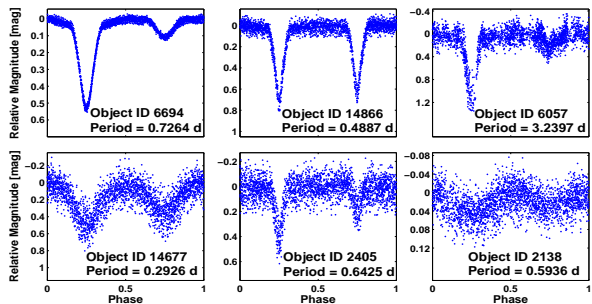,height=4.25cm,width=8cm}
\caption{Examples of eclipsing binaries detected by WHAT. }
\label{fig:lcs}
\end{center}
\end{figure}

To demonstrate the capability of WHAT in detecting small-amplitude periodic variables, we present in Figure \ref{fig:10290} one such case. Detection is highly significant, even though the signal's amplitude is only 4.2~mmag. In Figure \ref{fig:lcs} six eclipsing binaries of different types, detected in the same field, are presented.

We are currently engaged with finalizing our photometric pipeline and observing additional fields.

\acknowledgments{The WHAT project has been made possible by grants
from the Sackler Institute of Astronomy and from the Hungarian
Scientific Research Fund (OTKA M-041922, T-038437). Work of
G.~B. was supported by NASA through Hubble Fellowship grant
HF-01170.01 awarded by the STScI, which is operated by the AURA,
Inc., for NASA, under contract NAS 5-26555.}


\begin{references}
\reference Bakos, G., Noyes, R.~W., Kov{\' a}cs, G., Stanek, K.~Z., Sasselov, D.~D. \& Domsa, I.\ 2004, \pasp, 116, 266 
\end{references}
\end{document}